\documentclass[twocolumn,showpacs,preprintnumbers,superscriptaddress, amsmath,amssymb]{revtex4}

\pdfoutput=1

\usepackage{graphicx}
\usepackage{dcolumn}
\usepackage{bm}

\begin{document}

\preprint{}

\pdfoutput=1

\title{Current Fluctuations in Rough Superconducting Tunnel Junctions}

\author{Georg Heinrich}
\affiliation{Physics Department, Arnold Sommerfeld Center for Theoretical Physics and Center for NanoScience,\\  Ludwig-Maximilians-Universit\"{a}t, Theresienstrasse 37, 80333 Munich, Germany}%
\affiliation{
IQC and Department of Physics and Astronomy, University of Waterloo,\\
200 University Ave W, Waterloo, ON, N2L 3G1, Canada}%

\author{F. K. Wilhelm}
 \email{fwilhelm@iqc.ca}
\affiliation{
IQC and Department of Physics and Astronomy, University of Waterloo,\\
200 University Ave W, Waterloo, ON, N2L 3G1, Canada}%

\date{\today}

\begin{abstract}
Intrinsic noise is known to be ubiquitous in Josephson junctions.
We investigate a voltage biased superconducting tunnel junction including a very small number of pinholes - transport channels possessing a transmission coefficient close to unity.
Although few of these pinholes contribute very little to the conductance, they can dominate current fluctuations in the low-voltage regime. We show that
even fully transparent transport channels between superconductors contribute to shot noise due to the uncertainty in the number of Andreev cycles. 
We discuss shot noise enhancement by Multiple Andreev Reflection in such a junction and investigate whether pinholes might contribute as a microscopic mechanism of two-level current fluctuators. We discuss the 
connection of these results to the junction resonators observed in 
Josephson phase qubits. 
\end{abstract}

\maketitle


\section{\label{ChapIntroduction}Introduction}

Implementing qubits using superconducting circuits \cite{nat:ClarkeWilhelm:SupercondQubits, rmp:MakhScho:EngineeringWithJosephson} seems to be one of the most promising approaches to design a quantum computer.
Various implementation schemes have been developed, \cite{nat:FriePate:SuperposMacroState, sci:vdWal:SuperposMacroPresistent, sci:ChioNaka:CohQDynamicsFluxQB, nat:NakaPash:CoherentControlCooperPairBox, nat:PashYama:QOscillations2coupledChargeQB, prl:MartNam:RabiJosephsonJunctionQubit, sci:YuHan:CoherentOscQStatesJosephsonJunction, sci:McDeSimm:SimultaneousStateMeasurementCoupledPhaseQB}.
The crucial and indispensable device in all these setups is a Josephson tunnel junction.
Hence, microscopic understanding of this kind of junction and all possible details is essential to advance this field.
Current fluctuations in Josephson Junctions, as they are discussed in this paper, are of particular importance as they contribute to decoherence.

\subsection{\label{Intro_Decoherence}Decoherence, 1/f noise}
\label{Intro_1oFNoise}

One of the major challenges for the realization of practical quantum computing is to perform a sufficient number of quantum manipulations within the coherence time. The need to maintain quantum coherence during the operation is especially difficult to achieve in solid state systems which couple relatively strongly to uncontrollable environmental degrees of freedom, that generate quick decoherence.

After electromagnetic qubit environments have been successfully engineered to improve coherence, we are now mostly concerned with intrinsic noise of the solid state system.
The most prominent source of intrinsic decoherence is non-Gaussian 1/f noise \cite{apl:WelUrbCla:LowFrequencyNoise1K}, for which the spectral function behaves like $S(\omega) \propto 1/\omega$ \cite{rmp:DuttHorn:LowFrequ1oFNoise, rmp:Weissman:1oFNoiseKineticsInCondMat}.
1/f noise typically appears due to slowly moving defects in strongly disordered materials and is usually explained by an ensemble of two-level fluctuators coupling to the system under consideration. A heat bath causes uncorrelated switching events between the two states, which are described by a Poissonian distribution with mean switching time $\tau$. For a single fluctuator this leads to {\em random telegraph noise} (RTN). Superimposing several such fluctuators, using an appropriate mean switching time distribution $\rho(\tau)$, results in a 1/f noise spectrum. $1/f$ noise seriously limits
the operation of superconducting qubits \cite{Harlingen04,Muck05}.

Besides magnetic-flux fluctuations \cite{prl:KochDiVinCla:Model1/fFluxNoiseSQUID,preprint:FaoIof:MicroscOriginLowFreqFluxNoise}, critical-current fluctuations due to charge trapping at defects in the tunnel barrier \cite{prb:VanHRobe:DecoherenceCriticalCurrentFluctuation} is a prominent, possible mechanism for low-frequency 1/f noise in junctions of superconducting qubits.
As complement to this, we will investigate the intrinsic noise of Josephson tunnel junctions containing a few high-transmission channels, that potentially reside in the junction. We will address the question whether such defects might introduce another intrinsic source of current fluctuations leading to 1/f noise.

\subsection{\label{Intro_ModellJunction}Rough superconducting tunnel junctions}

\begin{figure}[!t]
\centering
  \includegraphics[scale=0.4]{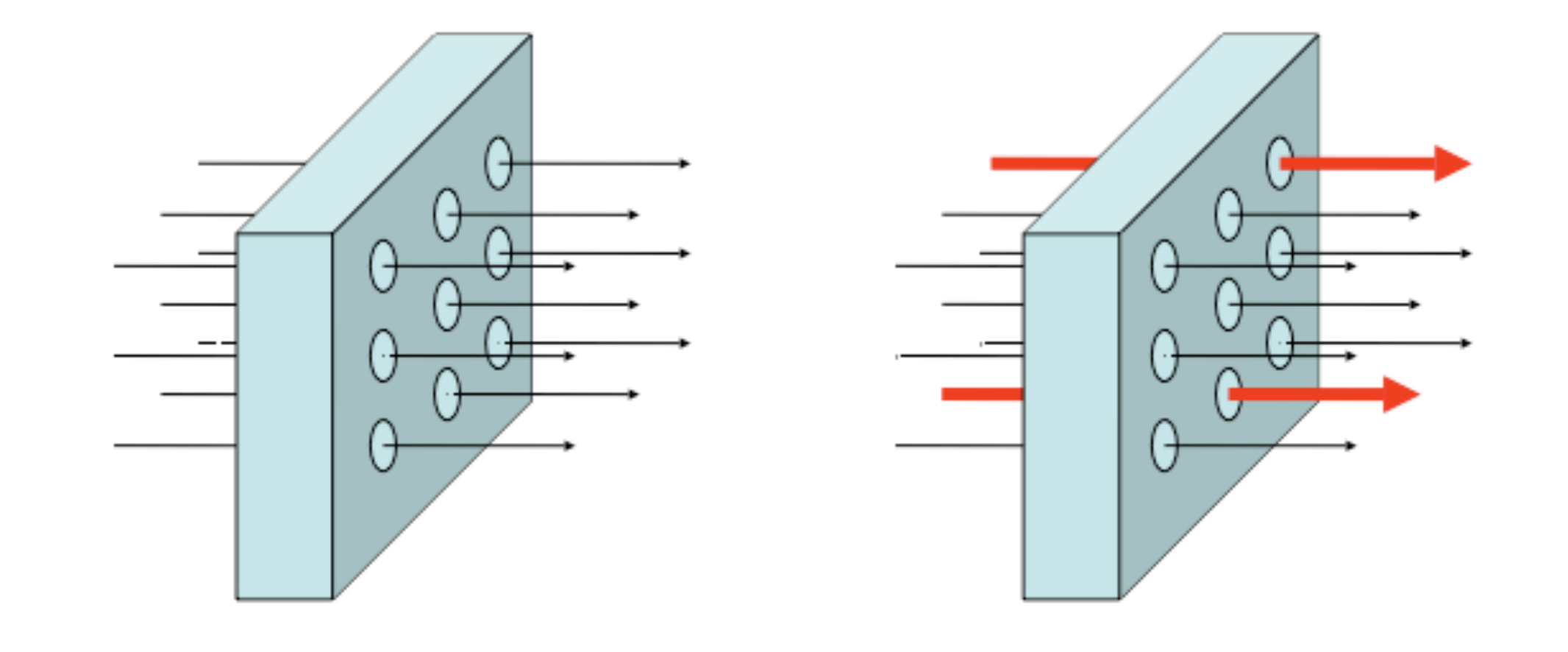}
  \caption[]{\label{fig:rough_tunnel_junction}Schematic diagram of the oxide layer of a Josephson Junction. Several transport channels are indicated. The arrow thickness corresponds to the respective transmission eigenvalue. Left: schematic diagram of an ideal Josephson tunnel junction. The transmission of all channels is small. Right: considered {\em rough Josephson tunnel junction}, i.e., a tunnel junction including some high-transmission channels, so-called {\em pinholes}.}
\end{figure}

A Josephson tunnel junction consists of two superconductors separated by a thin insulating oxide layer. Transport through such small scatterers can be described by quantum transport channels \cite{book:Datta:TransportMesoscopic} which, in our case, refer to the channels of the oxide.

In tunnel junctions, the transmission of all transport channels are assumed to be small. However, the fabrication process is usually not at all epitaxial quasi-equilibrium growth, thus one has to expect the oxide layer to be non-crystalline and disordered \cite{apl:EromsMooij:LowFreqNoiseInJJ, Supercond:OhMartPappas:LowLeakageSingCrystAlOBarrier}. 
We will take this into account by investigating {\em rough superconducting tunnel junctions}, where we assume that the junction additionally contains a few transport channels with very high transmission eigenvalues. {\em Pinholes}, see Fig.~\ref{fig:rough_tunnel_junction}, might occur as defects due to the fabrication process. Indeed, in Ref.~\cite{prl:DielBukk:AndRefEnhancedShotNoise}, the importance of pinholes was pointed out, but also Ref.~\cite{prb:ZarZimWilh:TwoChannelVSKondoMSET} discusses the significance of pinholes in mesoscopic devices, e.g., for the Kondo effect. 

There has been particular interest in pinholes to understand subharmonic gap structure in weak links.
In Ref.~\cite{prl:KleMil:ObservationMAR} the subharmonic gap structure of a tunnel junction was modeled by assuming that some channels have pinhole character. Analyzing superconducting qubits containing pinholes has additional motivations that are going to be reviewed in the following subsections.

\subsubsection{\label{introduction_MAR}Multiple Andreev Reflections (MAR)}

In Josephson Junctions with voltage bias smaller than the superconducting gap, direct tunneling of quasiparticles is impossible. In this case, charge transport is governed by {\em Multiple Andreev Reflection} (MAR), see Ref.~\cite{physica:BTK:MAR, prb:BlonTink:MAR, prb:OctTin:SubharmonicEnergy-GapStructure}.

{\em Andreev Reflections} occur at contacts between a scatterer and a superconductor.
In a system consisting of two superconductors separated by a scatterer there are two superconductor-scatterer interfaces where Andreev Reflection can take place. This leads to processes involving a sequence of Andreev cycles known as Multiple Andreev Reflection (MAR), in which charge can be transferred even for voltages smaller than $2\Delta/e$.
In general, an $n^{th}$ order MAR process, transferring $n$ charge quanta at a time, a so-called {\em Andreev cluster}, comprises $(n-1)$ Andreev Reflections and occurs above a threshold voltage $V_n=2\Delta/(en)$. For voltages below this MAR voltage $V_n$, the energy gap cannot be overcome by $(n-1)$ reflections. As these processes are composed of multiple transmission cycles, they sensitively depend on the electron transmission probability, i.e., the set of transmission eigenvalues characterizing the junction.

We can thus expect that rough superconducting tunnel junctions will be highly affected by MAR and we will see that even very few pinholes will have an extreme impact on the transport properties of the junction.

\subsubsection{\label{Intro_NoiseEnhancement}Noise enhancement due to MAR}

In Ref.~\cite{prl:DielBukk:AndRefEnhancedShotNoise}, shot noise~\cite{pr:BlaBue:ShotNoise} of $NbN/MgO/NbN$ superconductor-insulator-superconductor tunnel junctions was measured. The result 
shows noise enhanced by Andreev reflection.
The authors attribute this to the occurrence of MAR processes in pinholes in the MgO barrier. They model their data assuming Poissonian shot noise $2eI$, where they replaced the single charge quantum~$e$ by an effective transferred charge $q(V)$, due to MAR.

Such processes might be highly relevant as a source of intrinsic noise in superconducting qubit devices due to pinholes residing in the Josephson junction.
It has to be realized that, in the case of transport through pinholes, the shot noise is not governed by the simple Poisson formula which is only valid in the limit of small $T$.
The method, we will use, properly treats all possible transmission eigenvalues, and quantitative statements on the impact of rough barriers will be possible.

\subsubsection{Junction Resonators}
A new measurement revealing major intrinsic sources of decoherence in Josephson junction qubits was performed in Ref.~\cite{prl:SimLan:JunctionResonator}. The authors
observed characteristics of energy-level repulsion at certain frequencies, as predicted for coupled two-state systems. This structure of level-splittings 
was attributed to spurious resonators residing in the Josephson junction. Measurements of Rabi oscillations revealed that these resonators cause significant decoherence \cite{NatPhys:NeelMart:ProcessTomographyQMemory}. Similar to the scenario of charge trapping, mentioned before with respect to 1/f noise, the energy-level repulsion could be explained by assuming two-state current fluctuators in the junction.

Although other processes such as charge trapping within the junction barrier 
are believed to be relevant effects for realizing such spurious resonators, pinholes in rough tunnel junctions might be additional candidates for introducing two-state current fluctuators, see Section~\ref{FCS_PinholeJuncitonReso}.

The structure of this paper is as follows: After a short survey concerning the method used, we discuss leakage current of rough superconducting tunnel junctions. This is followed by a section regarding its noise properties. Finally we discuss the {\rm full counting statistics} of pinholes and whether they might contribute as a microscopic mechanism of two-level current fluctuators.


\section{Method}

We will be interested in the {\em full counting statistics} (FCS) of charge transfer through the junction, i.e., the probability distribution $P_{t_0}(N)$ for $N$ charge quanta to be transmitted within measurement time $t_0$ \cite{jmp:LeviLeeLeso:ElectronCountStat, epj:NazaKind:FCSofGeneralQMVariable, prl:PilgButt:StochPathIntFormFCS}. 
In addition to the noise characteristic, proportional to the second cumulant, this distribution will also supply us with higher, non-Gaussian cumulants such as they occur in RTN.

To calculate $P_{t_0}(N)$ we will apply the non-equilibrium Keldysh Green's function approach~\cite{AnnPhy:Naza:UniversalityWeakLocalization}. Within this scheme,
it is possible to employ several quantum field-theoretical methods, used in the transport theory of metals \cite{rmp:RammSmit:QuantumFieldInTransport}, and describe the system under consideration microscopically.
In the zero temperature limit, the cumulant generating function of a single mode voltage-biased Josephson Junction was calculated analytically in Ref.~\cite{prl:CueBel:FCSofMAR, prb:CueBel:dctransport}. We extend this result to multimode junctions containing $M$ transport channels which are characterized by a set of transmission eigenvalues $\{T_n\}$. In this case, based on~\cite{prl:CueBel:FCSofMAR, prb:CueBel:dctransport}, we immediately find for the cumulant generating function
\begin{eqnarray}
\label{eq:FullExpression}
S(\chi) & = & \frac{2t_0}{h} \sum_n \int_0^{eV} dE \nonumber \\
           & \times & \ln \left[ 1 + \sum_{n=0}^\infty P_n(E,V,T_n) (e^{in\chi}-1) \right],
\end{eqnarray}
where $P_n(E,V,T)$ is computed according to Ref.~\cite{prb:CueBel:dctransport}.

With this approach, we accurately take into account MAR within the rough 
junction model specified above. Thus we will find quantitative results going beyond an effective charge $q(V)$ alone.


\section{\label{chapter_leakage}Leakage Current}

We quantitatively investigate {\em leakage current}, i.e., current in the subgap voltage regime $eV < 2\Delta$ of a voltage-biased rough superconducting tunnel junction.
Using Eq.~(\ref{eq:FullExpression}), the average current of our junction, containing $M$ transport channels characterized by a set of transmission eigenvalues $\{T_n\}$ which is described by the distribution $\rho(T)$, is given by
\begin{equation}
\label{eq:Current}
I = \frac{2e}{h} M \int_{0}^{1} dT \rho (T) \int_{0}^{eV} dE \sum_{n} nP_n(E,V,T).
\end{equation}

\subsection{\label{HomoCont}Homogeneous multimode contacts}

For illustrative reasons we start from a homogeneous multimode contact between superconductors containing $M$ transport channels all with the same transmission eigenvalue $T_1$. The transmission eigenvalue distribution reads $\rho(T) = \protect{\delta(T-T_1)}$. In the case $M=1$, this would be a single-mode quantum point contact (QPC).
We compute conductance in units of the normal state conductance  $G_N = \frac{2e^2}{h} M \int dT \rho(T)$. From Eq.~(\ref{eq:Current}) it is clear that the normalized average current of a homogeneous multimode contact is the same as the one of a single-mode QPC which was already discussed in Ref.~\cite{prb:CueBel:dctransport}.

\begin{figure}[t]
 \centering
  \includegraphics[scale=0.27]{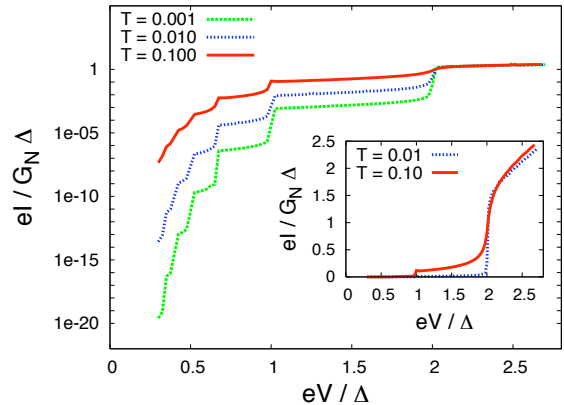}
  \caption[]{\label{C1Tsmall}Current through a homogeneous multimode contact with small transmission between superconductors as a function of bias voltage at $\protect{T=0K}$ on a logarithmic current scale. Inset: linear current scale.}
\end{figure}

\begin{figure}[b]
 \centering
  \includegraphics[scale=0.27]{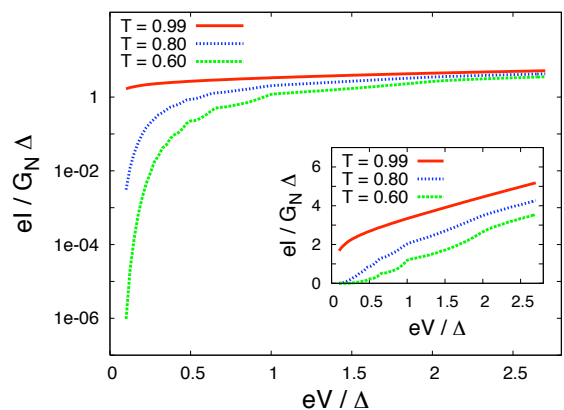}
  \caption[]{\label{C1Tlarge}Current through a homogeneous multimode contact with high transmission between superconductors as a function of bias voltage at $\protect{T=0K}$ on a logarithmic current scale. Inset: linear current scale.}
\end{figure}

For small transmission eigenvalues Fig.~\ref{C1Tsmall} shows the average current as a function of bias voltage for low transmission probability $T_1 \ll 1$ on a logarithmic and linear scale. We see that a contact with $T_1 = 0.1$ already develops a relatively large leakage current in the subgap regime $eV < 2\Delta$. Another immediate aspect, which will become important below, is its scaling. Conductance steps of size $T_1$ arise at MAR voltages $2\Delta/n$, demonstrating that the current is reduced by a factor of $T_1$  at each step.
For a single-mode QPC it was shown before that current transport for small transmission eigenvalues in the voltage interval $[2\Delta/(ne),2\Delta/(n-1)e]$ is dominated by the $n^{th}$-order MAR process.
In Ref.~\cite{prb:CueBel:dctransport}
the authors explicitly showed this $T^n$ dependence within a perturbative calculation.

For high transmission contacts we note that
perturbative approaches in $T$ will fail and it is necessary to use non-perturbative methods, as we do here. This will become in particular important for deriving quantitative results for rough junctions containing low- {\em and} high-transmission channels. Fig.~\ref{C1Tlarge} shows the current for a range of transmission probabilities $T\geq0.6$.
Especially at small voltages, the current through high-transmission modes is larger by orders of magnitude compared to the small transmission case.

\subsection{\label{Current_Rough}Rough tunnel juncitons}
We now turn to rough Josephson tunnel junctions assuming a small number of pinholes with transmission eigenvalues close to unity that reside in the junction.
We consider a contact with $M$ channels. A fraction $a$ of these channels has a high transmission eigenvalue $T_1$, the vast majority has a small value $T_2$, typical for tunnel contacts. Altogether, we consider the eigenvalue distribution
\begin{equation}
\label{transmission_dist_fct}
\rho(T) = a\delta(T-T_1) + (1-a)\delta(T-T_2),
\end{equation}
causing a normal conductance of
\[G_N = \frac{h}{2e^2}M[aT_1 + (1-a)T_2].\]
In Section~\ref{CharPin} we are going to discuss why this distribution captures the essential physics of more complicated distributions.
\begin{figure}[!htb]
 \centering
  \includegraphics[scale=0.27]{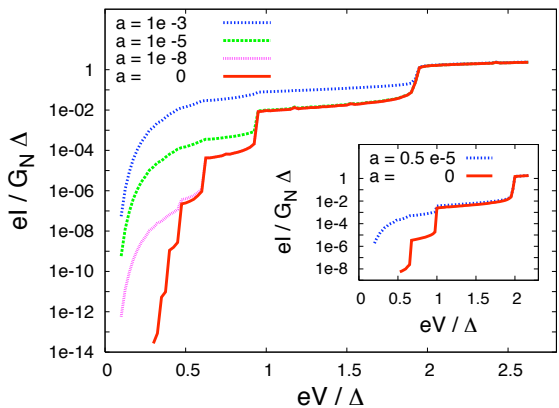}
  \caption[]{\label{mod_C1_Rough}Leakage current on a logarithmic scale as a function of bias voltage at $\protect{T=0K}$ for a rough superconducting tunnel junction with transmission eigenvalue distribution $\rho(T) = a\delta(T-T_1) + (1-a)\delta(T-T_2)$, $T_1=0.6$, and $T_2=0.01$. Inset: $T_1=0.6$, and $T_2=0.003$. The different curves refer to different pinhole fractions $a$ within all transport channels. }
\end{figure}

We calculate the current for this model, taking into account the very different transport properties of $T_1$ and $T_2$ channels. The result for two scenarios with extremely small pinhole fraction $a$ is shown in Fig.~\ref{mod_C1_Rough}.
We see that, starting at high voltages, the current follows the well-known subharmonic gap structure curve for tunnel transmission coefficient $T_2$ only above a certain voltage, depending on $a$. 
However, below that voltage, due to the fact that the current carried by tunnel transmission channels is reduced by a factor of $T_2$ each time the voltage passes another MAR voltage $2\Delta/n$, the highly transmissive channels dominate, leading to a smooth, weakly structured subgap contribution that level off into a plateau before it drops again.
Consequently, at sufficiently low voltages, the current through rough tunnel contacts is overwhelmingly carried by the pinhole fraction.

With respect to the pinhole fraction $a$, our quantitative treatment demonstrates that in tunnel junctions we can only have very few pinholes, which Refs. \cite{prl:KleMil:ObservationMAR} seems to overestimate. From Fig.~\ref{mod_C1_Rough}, for example, we infer that in a junction, where a possible measurement shows two full current steps each scaling with a factor of $T=0.01$ at $eV=2\Delta$ and $eV=2\Delta/2$, respectively, we can have roughly less than 1 out of $10^6$ channels with transmission $T\geq0.6$!

\subsection{\label{CharPin}Characterizing Pinhole Thresholds}
As we have seen, below a certain voltage, a high transmission channel residing in a rough Josephson tunnel junction dominates high-order subharmonic steps in the current characteristic. We can use this result to characterize the fraction of pinholes in all transmission channels by very sensitive current measurements.

In Ref.~\cite{ieee:LanNam:Banishing} a current-voltage plot for an $Al$-$Al_2O_3$-$Al$ junction used in a Josephson-junction qubit is presented. At $eV=2\Delta$ the measured current shows a subharmonic step corresponding to a tunnel transmission eigenvalue of $T=0.003$. The second current drop at $eV=\Delta$ is indicated but the measurement does not resolve the next expected plateau. The experimental result is consistent with the calculation presented in the inset of Fig.~\ref{mod_C1_Rough} for a pinhole fraction $a=0.5 \cdot 10^{-5}$. This corresponds to one pinhole of $T_1=0.6$ in a junction of $1/a=200\, 000$ channels. Actually, in Ref~\cite{prl:SimLan:JunctionResonator}, the number of transport channels for the junction under consideration~\cite{ieee:LanNam:Banishing} is estimated to this order of magnitude, indicating that the existence of pinholes in state-of-the-art superconducting qubit devices is compatible with current measurements.

Indeed, new design concepts have lead to a significant reduction of the junction size (see Ref.~\cite{prl:StefAnsm:StateTomography}) and with it a supression of intrinsic noise. 
Clearer insight would be provided by highly sensitive current-voltage measurements at voltages stretching out over several current steps at $V_n=2\Delta/n$.
In the following, we will assume a small number of pinholes.

Finally, this provides the justification for the very simple transmission distribution function~(\ref{transmission_dist_fct}).
The transmission eigenvalues are determined by WKB, $T = \exp (-\kappa d)$, and this way, depend on the junction width $d$. Then the pinhole transmission eigenvalues might be related to a distribution if widths $\rho (d)$ of the oxide layer separating the superconductors. Considering the strict non-negativity of $d$, a lognormal distribution might be appropriate for describing $\rho(d)$ for the pinholes.

All this can be done in our approach, but as we have seen above, in state-of-the-art superconducting qubit devices we might only have a small, single-digit number of pinholes in a huge junction. Thus, doing statistics is not necessary and considering a single value $T_1$ to represent the pinhole transmission eigenvalue distribution, as done in Eq.~(\ref{transmission_dist_fct}), is a sufficient way to take them into account.


\section{\label{ChapNoise}Noise}

We will examine the noise properties of rough superconducting tunnel junctions. As we have seen a small amount of pinholes contributes very little to the conductance in the low-voltage regime. Now we are going to show, that pinholes on the contrary do dominate current fluctuations.

Using the cumulant generating function~(\ref{eq:FullExpression}), we find for the zero-frequency noise
\begin{equation}
S_I=\frac{4e^2}{h}\int_{0}^{eV} dE [ \sum_{n} n^2P_n(E,V,T) - (\sum_{n}nP_n(E,V,T))^2 ].
\label{S_I}
\end{equation}

\subsection{\label{NoiseHomo}Homogenous multimode contacts}

Again, as in Section~\ref{chapter_leakage}, we start from homogeneous multimode contacts with each channel having the same transmission eigenvalue and
begin with small transmission $\protect{T\ll1}$.
\begin{figure}[!t]
 \centering
  \includegraphics[scale=0.27]{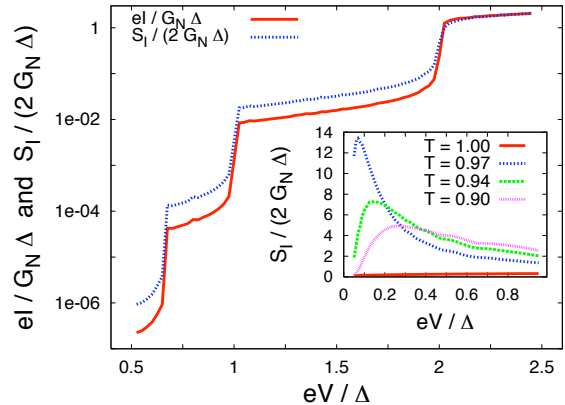}
  \caption[]{\label{fig:C1C2Tlow}Noise and current as a function of bias voltage for a homogeneous superconducting tunnel junctions with transmission eigenvalue $T=0.01$. Inset: noise characteristics for junctions with large transmission.}
\end{figure}
Fig.~\ref{fig:C1C2Tlow} shows noise and current characteristics.
In this case, as there is always one very dominant MAR-process, which causes charge transport to be dominated by quanta of $q(V)=e(1+\lfloor 2\Delta/eV \rfloor)$, the noise scales with this additional charge factor. Thus, in the small-transmission regime Poissonian shot noise $S_I=2eI$ with modified charge quantum $nq(V)$, properly explains the observed noise features.

In the case of large-transmission eigenvalues, inset of Fig.~\ref{fig:C1C2Tlow}, the noise characteristic changes dramatically.
For very high probabilities $T$, the noise increases with decreasing voltage in the subgap regime. Depending on the value of $T$, it develops a maximum, but falls off again at even lower V. Remarkably, and in strong contrast to any simpler model, we note that a contact with perfect transmission $T=1$ shows low, but finite noise. This is markedly different from the normal conducting case where, given the shot noise formula $S_I = (Ve^3/\pi \hbar)T(1-T)$, we would anticipate zero noise in the case of perfect transmission. Furthermore, we see that the larger the transmission the steeper and higher is the noise ascent for small voltages. For high $eV$, the high transmission curves approach the $T=1$ characteristic. Thus, altogether we see that in this case the description with pure Poissonian shot noise with modified charge quantum is insufficient and the generalization used in the rest of this paper shows new features.

\begin{figure}[!t]
 \centering
  \includegraphics[scale=0.27]{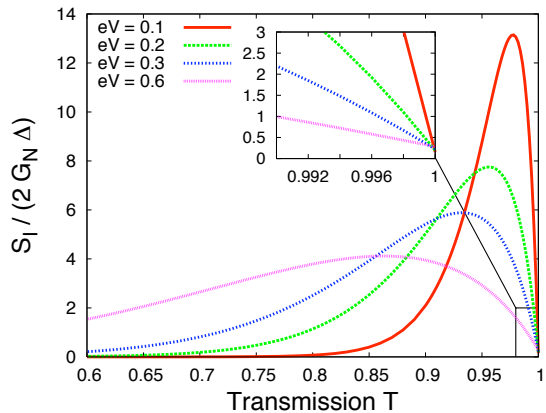}
  \caption[]{\label{fig:C2_T_2D}Noise $S_I$ as a function of transmission eigenvalue $T$ for a homogeneous superconducting tunnel junction where the bias voltage is set as parameter. Inset: enlargement around $T=1$.}
\end{figure}

It is instructive to look at the noise curve from a different perspective. Focussing on the $T$-dependence, in Fig.~\ref{fig:C2_T_2D}, we set voltage as a parameter and plot noise as a function of transmission.
The noise develops a maximum at high transmission values. As noticed before, each curve falls off to a finite residual noise level at $T=1$. For smaller voltages, the maximum becomes more and more pronounced and it seems to be squeezed into the high-transmission regime.
On the order of $eV=0.1$ only channels with very high transmission significantly contribute to the noise. 

The explanation of the noise features for high transmission is more involved.
From Eq.~(\ref{S_I}) we  see that the noise can be expressed in terms of the variance of $P_n(E,V,T)$, which is the probability for a MAR-process transferring charge quanta $ne$
\begin{equation}
S_I=\frac{4e^2}{h}\int_{0}^{eV} dE [ \langle n^2 \rangle - \langle n \rangle^2 ] = \frac{4e^2}{h}\int_{0}^{eV} dE \; {\rm Var}(n).
\label{S_I_Var}
\end{equation}
Thinking of shot noise as partition noise, for a single-mode normal conductor with perfect transmission $T=1$, there is no uncertainty whether a particle is transmitted or reflected. We find zero noise. In the superconducting case, due to perfect transmission, we are still certain about charge transfer taking place, but an additional uncertainty is introduced.
For high transmission including $T=1$, there are many different MAR-processes contributing to charge transport, which is described by the probability distribution $P_n$. This additional uncertainty is the qualitative physical explanation of the finite noise observed in the case of perfect transmission.

\subsubsection{Toy Model}

To clarify the essential physics, referring to the full computer-algebraic calculation is unsatisfactory. Thus, we will try to explain the basic noise features with the use of the toy model presented before in Ref.~\cite{prb:CueBel:dctransport}. Originally, this model was introduced to illustrate how to calculate the cumulant generating function of a weak link with voltage bias in an easy, analytically solvable case. We summarize the basic simplifying assumptions.

We only look at voltages equal to one of the MAR voltages $eV=2\Delta/n$, and for each of them we only take into account one MAR process, namely the most relevant one which transfers
\begin{equation}
\label{ExplainWithToyN_eV}
N= \bigg\lfloor \frac{2\Delta}{eV} \bigg\rfloor +1
\end{equation}
charge quanta. This simplifies the cumulant generating function $S(\chi)$ to the one of a binomial distribution.
Furthermore, in this model, Andreev reflection above the gap is neglected and the Green's function is simplified by assuming a constant density of states above the gap.

\begin{figure}[!b]
 \centering
  \includegraphics[scale=0.27]{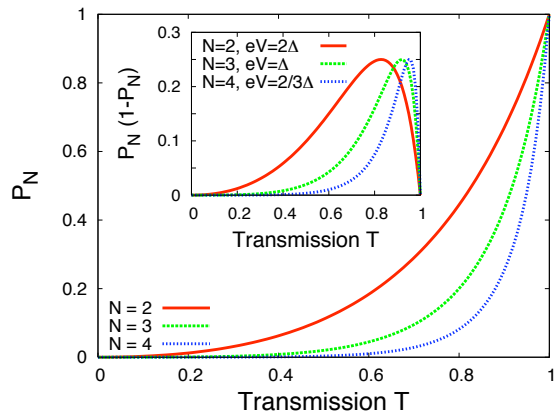}
  \caption[]{\label{ToyModelP_NundC2}Toy model probability $P_N$ to transfer $N$ charges at a time as a function of transmission eigenvalue $T$. Inset: expression $P_N(1-P_N)$ that occurs as a term in the $2^{nd}$ cumulant (Eq.~(\ref{C2inToy})) as a function of transmission eigenvalue $T$.  }
\end{figure}

The cumulant generating function for the toy model reads
\begin{equation}
\label{ExplainWithToyS}
S(\chi) = \frac{2eVt_0}{h} \ln \left[1+P_N (e^{iN\chi}-1) \right].
\end{equation}
For every MAR voltage $V$, another specific transport process with probability $P_N$ is relevant, Eq.~(\ref{ExplainWithToyN_eV}). We emphasize that, due to this, the argument of the logarithm in Eq.~(\ref{ExplainWithToyS}) depends on voltage via the selection of the relevant $P_N$.
Fig.~\ref{ToyModelP_NundC2} shows the toy-model probabilities $P_N$ as a function of transmission eigenvalue $T$. For perfect transmission, as we have reduced the system to a binomial distribution involving only a single transport process, each probability is unity. The probabilities for $N \geq 2$ and imperfect transmission are always smaller than in the normal conducting case, because a higher-order process is necessary in order to transfer charge. For large charge quanta $N$, very high transmission is necessary, since many Andreev reflections are involved in such a process. Thus, for larger $N$, i.e., small voltage bias (Eq.~(\ref{ExplainWithToyN_eV})), nonzero probabilities are more and more shifted to the high-transmission regime.

The second cumulant
\begin{equation}
\label{C2inToy}
C_2 = N^2 \frac{2eVt_0}{h}P_N(1-P_N)
\end{equation}
is proportional to the noise correlator. The expression $P_N(1-P_N)$, which matches the one in the traditional shot-noise formula if we replace $T$ by $P_N$, is displayed in the inset of Fig.~\ref{ToyModelP_NundC2}. For large N, i.e., small voltage, the maximum is shifted and squeezed into the high-transmission regime.

So, altogether, we distinguish two mathematical ingredients to the noise. One is the expression $P_N(1-P_N)$ that we just discussed. Additionally, there is the prefactor $N^2(2eVt_0/h)$.
In the small voltage regime it results in noise enhancement that behaves approximately like $1/V$.
As the noise is determined by the product of both parts, for a fixed transmission coefficient, there will be a voltage regime where the noise gets enhanced by lowering the applied voltage bias. However, at some voltage, or conversely for some $N$, $P_N(1-P_N)$ will overcompensate this increase and reduce the noise again.
To summarize, the toy model still explains noise enhancement by an increased charge quantum. The decrease of noise at very low voltage follows from the overcompensation of this effect by the decrease of transfer probability in the expression $P_N(1-P_N)$.
\begin{figure}[!t]
 \centering
  \includegraphics[scale=0.27]{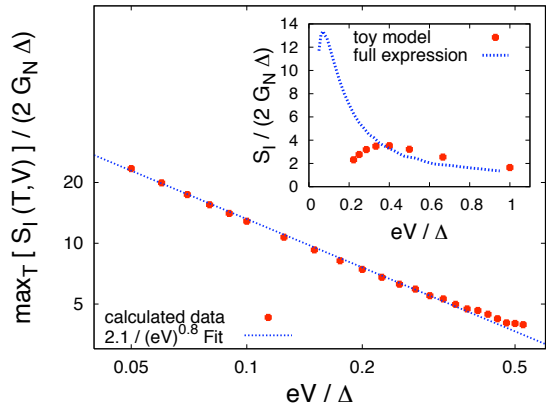}
  \caption[]{\label{ToyModel_Toy_and_full}Maximal noise $\max_T \left[S_I(T,V)\right]$, optimized with transmission as a parameter, as a function of bias voltage on a double logarithmic scale for a homogeneous contact between superconductors. The calculated data using the full expression, Eq.~(\ref{eq:FullExpression}), are fitted using the given power law. Inset: Comparison of the noise results calculated for a $T=0.97$ homogeneous contact between superconductors as a function of bias voltage using either the full expression or the toy model.}
\end{figure}

In the inset of Fig.~\ref{ToyModel_Toy_and_full}, for comparison, the noise calculated using the full expression and the toy model at MAR voltages, is presented in a single plot. The simplified model qualitatively shows the basic features of our numerical calculation. Nevertheless, there is a huge quantitative difference. Thus, we realize that the toy model is qualitatively sufficient but it fails dramatically to provide quantitative results. Thus, for quantitative calculations, the extensive calculation used in the remainder of this paper is essential.

\subsubsection{What is driving the noise increase?}

We can ask the question: what is the maximal noise at a given voltage? This means, for fixed voltage bias, we use the transmission eigenvalue as a parameter to find the maximal value.
In the toy model, Fig.~\ref{ToyModelP_NundC2}, $\max_T \left[P_N(1-P_N)\right]$ is always $1/4$. Thus, here, the maximal noise $\max_T \left[S_I(T,V)\right]$ depends only on the prefactor in Eq.~(\ref{C2inToy}). Consequently, for small voltages, it approximately scales like $1/V$.

For the full theory, in Fig.~\ref{ToyModel_Toy_and_full} the maximal noise $\max_T \left[S_I(T,V)\right]$ is plotted against voltage bias on a double logarithmic scale. In the small-voltage regime, the data can be fitted well using a power-law. We find
\[ \max_T \left[S_I(T,V)\right] \propto \frac{1}{V^{0.8}}.\]
Thus, although quantitative statements resulting from the toy model and from the full expression differ significantly, we see that the maximal noise at given voltage follows a similar power law with an exponent of 0.8 instead of unity. Hence, even in the much more complicated situation, including multiple MAR processes, the inherent $1/V$ dependence, which basically results from the increased charge quanta, seems to play a major role.

\subsection{Noise of rough superconducting tunnel junctions}

\begin{figure}[!b]
 \centering
  \includegraphics[scale=0.27]{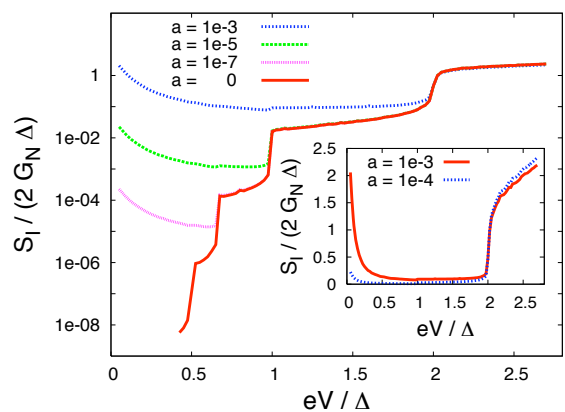}
  \caption[]{\label{fig:C2T0_986T0_01}Noise $S_I$ on a logarithmic scale as a function of bias voltage for a rough superconducting tunnel junction characterized by the transmission eigenvalue distribution $\rho(T) = a\delta(T-T_1) + (1-a)\delta(T-T_2)$, $T_1=0.986$ and $T_2=0.01$. The different curves refer to different pinhole fractions $a$ within all transport channels. Inset: linear noise scale.}
\end{figure}

We now return to the model of Section~\ref{CharPin}. There, we looked at a rough superconducting tunnel junction with eigenvalue distribution
given by Eq.~(\ref{transmission_dist_fct}). Here, we are concerned with the noise generated in this kind of junction. Fig.~\ref{fig:C2T0_986T0_01} shows the result.

In contrast to a normal tunnel junction we see a dramatic change in the noise characteristic due to very few pinholes with an enormous noise increase at small voltages. As in the case of leakage current, at a certain point in the subgap regime, the pinholes begin to dominate the noise characteristic. In this range, $S_I$ is solely carried by the few pinholes.

Together with our results in Section~\ref{chapter_leakage} this demonstrates one of our central results: although a small amount of pinholes residing in the junction contributes very little conductance, it can dominate current fluctuations in the low-voltage regime.
As pointed out before, sensitive measurements of the leakage current will provide an estimate on the amount of pinholes that might be contained in the considered junction.

The considered pinhole transmission eigenvalue of $T=0.986$ is chosen in order
to display all structure at voltages down to $eV=0.05\, \Delta$. Nevertheless, analogous to Section~\ref{NoiseHomo}, we can add two more aspects: Firstly, for smaller voltages than resolved in Fig.~\ref{fig:C2T0_986T0_01}, the noise will show a maximum and then will fall off again. Secondly, considering higher values of transmission will lead to an even steeper and higher ascent, starting at smaller voltages.


\section{\label{FCS}Full Counting Statistics of Pinholes}

We will investigate the {\em full counting statistics} (FCS) of charge transport through pinholes of a rough superconducting tunnel junction.
Given the definition of the cumulant generating function, the FCS is calculated by Fourier transformation.
Using this insight into the transport process, we will furthermore discuss a possible model of high transmission channels as microscopic origin of two-level current fluctuators.

\subsection{Resolution of structure in the Full Counting Statistics}

To determine the probability distribution $P_{t_{0}}(N)$, we have to set the measurement time $t_0$.
In general, the calculation of the cumulant generating function for a voltage biased Josephson junction is significantly complicated due to the ac Josephson effect~\cite{prl:CueBel:FCSofMAR, prb:CueBel:dctransport}.
In order to make computation feasible, and to avoid interpretation difficulties of arising 'negative probabilities' in the superconducting system~\cite{prl:Bel:Naz:FCS_e_transfer_super},
$t_0$ must be sufficiently longer than the inverse of the Josephson frequency $T_{\rm J}=h/2eV$. (See Ref.~\cite{prb:CueBel:dctransport} for further details).
Consequently $T_{\rm J}$ sets a time scale in our approach and there is a lower bound for the measurement time $t_0$.

We consider a contact with transmission eigenvalue $T=0.936$ at low bias voltage $eV=0.3 \Delta$, where qubits might be operated, and take into account two different measurement times $t_0=10 \; T_{\rm J}$ and $t_0=100 \; T_{\rm J}$. Fig.~\ref{mod_FCS_eV0_3_t10_t100} shows the results:
\begin{figure}[!t]
 \centering
  \includegraphics[scale=0.27]{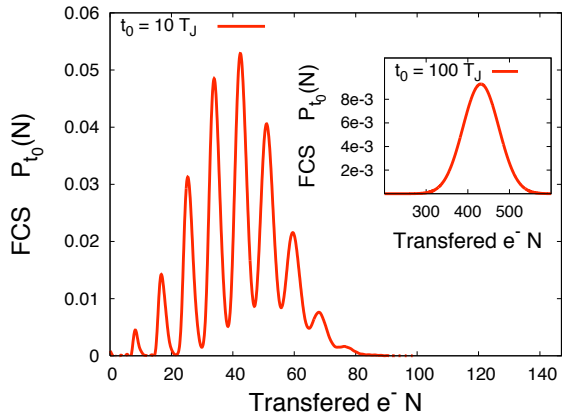}
  \caption[]{\label{mod_FCS_eV0_3_t10_t100}FCS for a transport channel between superconductors with transmission $T=0.936$ and voltage bias $eV=0.3\Delta$ at measurement time $t_0=10\; T_{J}$. Inset: measurement time $t_0=100\; T_{J}$}
\end{figure}
For the long measurement time the FCS is Gaussian.
In contrast, for $t_0=10\; T_{J}$ we see a rich comb structure.

We will discuss this comb structure and its origin in detail later on. Here, we want to point out that this structure turns into a Gaussian for long measurement time $t_0$. This is as we would expect: if we sum the number of transfered charges over a very long time it will become possible, instead of considering individual MAR processes with their specific probabilities, to just assign an average likelyhood for one elementary charge quantum to be transfered. Thus, in the long measurement time limit, transport can be described by a sum of many independent and identically-distributed events what results in a Gaussian. This is the essence of the central limit theorem used in statistical physics. Indeed, the problem above can be related to the {\em quasi-ergodic hypothesis}. Hence, it is clear that for very long measurement times the comb structure, due to individual, discrete transport processes, is washed out.

For significantly higher voltages, like $eV=1.5\Delta$, the most relevant transport processes transfer much smaller charge quanta. It turns out that consequently, in this case, discrete structures in the FCS cannot be resolved using the time interval $t_0=10\; T_{\rm J}$. Despite these limitations concerning $t_0$, we can resolve structure in the FCS for a limited parameter window.

\subsection{\label{FCS_PinholeJuncitonReso}Pinholes as Junction Resonators?}

\subsubsection{Motivation}

We are now coming back to the questions whether pinholes might explain decoherence from junction resonators in phase qubits or 1/f noise~\cite{rmp:DuttHorn:LowFrequ1oFNoise, rmp:Weissman:1oFNoiseKineticsInCondMat, prb:VanHRobe:DecoherenceCriticalCurrentFluctuation}.
Thinking of the different possible MAR processes, which transfer different sizes of charge quanta, a pinhole might introduce current fluctuators: imagine a high transmission channel, i.e., a pinhole hidden in the junction. Two different MAR processes $A$ and $B$ transfer charge in two different quanta $n_Ae$ and $n_Be$. Thus, we might think of two current states $|A\rangle$ and $|B\rangle$; each of them carry charge using only one of the distinct MAR processes $A$, $B$, respectively. Due to the differently sized Andreev clusters being transfered, the two states will cause two different currents. In principle the mechanism is similar to the idea of charge-trapping~\cite{prb:VanHRobe:DecoherenceCriticalCurrentFluctuation}, where a trapped charge blocks tunneling through a transport channel. There, one introduces an untrapped state $|\tau_u\rangle$ causing high current and a trapped state $|\tau_t\rangle$ causing low current. In comparison, we consider two current states $|A\rangle$ and $|B\rangle$ corresponding to charge transport by two different MAR processes and thereby causing two distinct currents.

\subsubsection{Calculation}
We invstigate whether this scenario results from a pinhole model. If this 
was the case, we would expect to find two distinct peaks in the FCS, where the first one refers to charge transport due to MAR process $A$ and the second one corresponds to MAR process $B$, each within the time interval $t_0$. Hence, let us see whether we find parameters that result in such an FCS.

\begin{figure}[!b]
 \centering
  \includegraphics[scale=0.27]{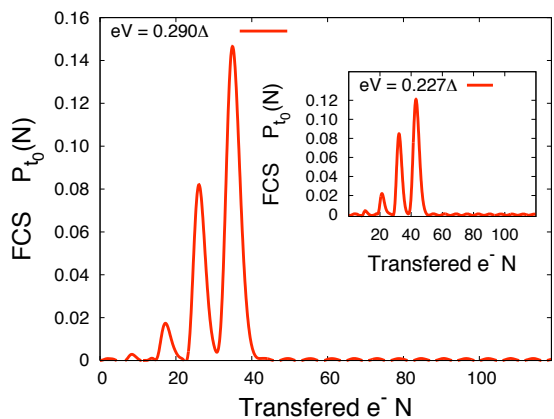}
  \caption[]{\label{FCS_2peaks}FCS for a transport channel between superconductors with transmission eigenvalue $T=0.99$ and voltage bias $eV=0.290\Delta$. Insert: $eV=0.227\Delta$. The measurement time is $t_0=4 \; T_{J}$.}
\end{figure}

We consider very high transmission channels, for instance $T=0.99$ and calculate the FCS for this transmission eigenvalue at two subgap voltages. The results are shown in Fig.~\ref{FCS_2peaks}. We find two very pronounced peaks in the FCS. Note that here the measurement time is very short but, despite some artifacts in the diagrams, the distribution still has a normalization close to unity.

\subsubsection{Attempted interpretation in terms of two-level fluctuator}

Given these pronounced peaks, does this result indicate a scenario where a pinhole via its different MAR processes might actually introduce a two-level current fluctuator? If we assume so, we associate the first peak with the case where charge transport is carried by MAR process $A$, i.e. charge transport in quanta of $n_Ae$ only.
Accordingly the second peak refers to the case where transport takes places via MAR process $B$, using charge quanta $n_Be$.

Taking a closer look at Fig.~\ref{FCS_2peaks} reveals a sharp boundary for the appearance of peaks towards large total charge numbers $N$.
In contrast, to the left, i.e. towards smaller $N$, we see small peaks next to the dominating ones. In fact, for a given voltage, there is a {\em lower} threshold for the MAR order, i.e., a {\em lower} bound on the minimal charge cluster being transferred in a single MAR process.
Furthermore, in addition to the dominant processes $A$ and $B$, there will also be finite probability for MAR of higher order, i.e., current flow via even larger quanta than $n_Ae$ or $n_Be$. So according to the two-level interpretation, identifying each peak with charge transport due to different MAR processes, we would expect this boundary to be reversed, namely: a sharp boundary for the existence of peaks towards small $N$, due to the lower bound on the charge cluster size, and additionally, little peaks towards large $N$, due to the finite probability for MAR of higher order than the two dominant ones, $A$ and $B$.

The second aspect is the spacing between the peaks. For the distributions in Fig.~\ref{FCS_2peaks}, the distance is slightly larger than the smallest possible charge quantum $\protect{(\lfloor 2\Delta/eV \rfloor +1)}$, i.e., for the main panel 7 (9 for the inset). With respect to the MAR threshold, this is roughly the size of the average charge quantum that we would expect to be transferred by a single Andreev cluster.
In Fig.~\ref{FCS_2peaks}, from the number of transferred charges and the minimal Andreev cluster size, we infer that, within the measurement time $t_0$, roughly 5 MAR processes contributed to the rightmost peaks.
In the above two-level scenario, $A$ and $B$ are adjacent MAR processes meaning their transferred charge quanta differ only in one elementary charge. Thus, if a pinhole introduced a two-level current fluctuator where each peak refers to current flow via distinct MAR processes $A$ and $B$, in Figure~\ref{FCS_2peaks} we would expect a peak spacing of $\Delta N = 5$ rather than a value larger than $(\lfloor 2\Delta/eV \rfloor +1)$. This makes the two-level fluctuator hypothesis inconsistent.

\subsubsection{Alternative, consistent interpretation}
Thus, the structure we have seen in the FCS of a pinhole does not correspond to the scenario of a two-level current fluctuator as suggested above. In fact, the description of the probability distribution becomes consistent if we identify each peak with the number of attempts being successful to transmit an Andreev cluster: within the measurement time $t_0$ we might think of a total number of attempts to transfer charge cluster, where the actual size of the quantum might differ due to the individual, possible MAR processes. In the distributions of Fig.~\ref{FCS_2peaks}, each rightmost peak corresponds to the case where every attempt is successful to transfer an Andreev cluster, so we get the sharp boundary observed for the appearance of peaks towards large $N$. The next peak to the left corresponds to the case where exactly one attempt fails and so on. Thus, the peaks are naturally separated by a distance larger than $(\lfloor 2\Delta/eV \rfloor+1)$, namely the average Andreev cluster size transferred in case of a successful attempt. As the actual size of successfully transmitted clusters might differ due to the individual MAR processes, the pronounced peaks are broadened. The comb structure in Fig.~\ref{mod_FCS_eV0_3_t10_t100} can be explained the same way. Here, in contrast to Fig.~\ref{FCS_2peaks}, due to smaller transmission, the case where every attempt is successful is not the most likely one.

\subsubsection{\label{Chap_FCS_PinReso_Conclusion}Conclusion}
To summarize this section, we have discussed the possibility of a pinhole to introduce a two-state current fluctuator due to its different MAR transport processes. This is conceptually similar to the mechanism of charge-trapping, Ref.~\cite{prb:VanHRobe:DecoherenceCriticalCurrentFluctuation}. Although at first sight it is suggestive to relate the observed peak structure to distinct MAR processes, a more detailed analysis suggests a very different but consistent interpretation in terms of successful transport attempts of Andreev cluster. Taking this into account, we see no clear evidence that a pinhole might be a microscopic origin for introducing two-level current fluctuators. Charge-trapping in junctions is probably one of the most relevant mechanisms. However, it might be in particular interesting to think about such a process opening and closing a very high transmission channel i.e. a pinhole. Due to the large charge quanta being transfered, the process of trapping and untrapping might result in high magnitudes of current fluctuations.
This picture may change if electron-electron interaction is included, given that the large charge quanta in a pinhole may efficiently block large parts of the junction.

A very intuitive picture might be an occupied upper Andreev bound state~\cite{cond-mat:Mart:SupercondQBPhysicsJosephsonJunc}, that causes a repulsion within the channel. Nevertheless, in the case of voltage bias, such a state with energy $E_{J} = \Delta[1-T \sin^2(\phi(t)/2)]^{1/2}$, where $\phi(t)$ is the superconducting phase, might be adiabatically carried above the gap within the actual Josephson cycle directly after population. Further research might clarify this scenario.


\section{Conclusion}

We have investigated voltage-biased {\em rough superconducting tunnel junctions} containing some high transmission channels, {\em pinholes}. We have accomplished this using the method of full counting statistics formulated within the non-equilibrium Keldysh Green's functions technique. Based on this microscopic approach, we were able to properly quantify physical effects due to low- {\em and high-} transmission channels in a single junction.

By exploring {\em leakage current} of such systems, we observed that a tunnel junction may contain much fewer pinholes than previously speculated~\cite{prl:KleMil:ObservationMAR}.
We further demonstrated how highly sensitive current measurements can clarify the existence of pinholes. We pointed out that existing current measurements done for junctions of the superconducting qubit devices~\cite{prl:SimLan:JunctionResonator}, do not strictly rule out the existence of a hidden pinhole.

Furthermore, we examined {\em noise} properties.
We demonstrated that even very few pinholes give rise to a drastic increase of the noise in the very low subgap voltage regime. Thus, although few pinholes might contribute very little to the average current, they can dominate current fluctuations.
Although details of this noise enhancement, comprising contributions of several MAR processes, turned out to be quite complicated, we proposed that the physical essence of the observed noise boost still lies in the increased charge quantum that is transfered.
To do this, we compared the explicit noise calculation to a simplified model. This showed qualitative agreements, and thus illuminated some essential features, but failed quantitatively, therefore demonstrating the need of a full calculation.

Finally, we investigated the {\em full counting statistics} (FCS) of charge transport through pinholes. Despite limitations concerning the measurement time $t_0$, we could resolve non-Gaussian structure in the FCS for a limited parameter window.
We discussed a possible model of high-transmission channels as a microscopic origin of two-level current fluctuators.
Indeed, for certain voltage parameters, the FCS shows a two-level peak structure. From a more detailed analysis we inferred that this structure {\em cannot} be related to charge transport by distinct MAR processes. Thus, given the dc part of the probability distribution, we find no evidence that a pinhole might introduce an additional source of two-level current fluctuators. We presented an alternative, consistent interpretation of the observed peak structure in terms of successful transmission attempts of Andreev clusters.

So far, our approach is limited to the stationary or quasi-stationary case. Improvements on this might incorporate time-dependence into the Keldysh Green's function approach. This may permit a more rigorous discussion of finite-frequency noise with respect to pinholes. Recently, first steps towards the discussion of time-dependence using this method have been made~\cite{prl:VaneNazaBelz:TransferVoltageDrivenQPC}.
Also, electron-electron interactions describing the traditional 1/f noise scenario for Josephson junctions should be included.

We acknowledge stimulating discussion with John M. Martinis, W. Belzig and Yu. V. Nazarov. This work was financially supported by NSERC through a discovery grant and QuantumWorks, EuroSQIP and Studienstiftung des deutschen Volkes.

\newpage 
\bibliography{literatur}

\end{document}